\documentclass[prl,aps,epsf,amsfonts,floats,twocolumn,amssymb,amsmath,groupedaddress,showpacs,floatfix,nofootinbib]{revtex4-2}

\usepackage{graphicx}
\usepackage{braket}
\usepackage[]{latexsym}
\usepackage{bm}
\usepackage{slashed}

\def\1{{_{1}}}\def\2{{_{2}}}

\def\noHe0{:\;\!\!\;\!\!:H_e(0):\;\!\!\;\!\!:}
\def\noHm0{:\;\!\!\;\!\!:H_\mu(0):\;\!\!\;\!\!:}

\def\1{{_{1}}}\def\2{{_{2}}}

\begin{document}
\title{Phenomenological implications of nonlocal quantum electrodynamics }
\author { Antonio Capolupo, Aniello Quaranta, Raoul Serao}
\affiliation{Dipartimento di Fisica "E.R. Caianiello" Universit\'a di Salerno,  and INFN - Gruppo Collegato di Salerno, Via Giovanni Paolo II, 132, 84084 Fisciano (SA), Italy }

\begin{abstract}

We analyze several phenomenological implications of a nonlocal generalization of quantum electrodynamics (QED). We compute the nonlocal corrections to the photon propagator up to one loop, and we show that nonlocality leads to a change of the Coulomb potential. We then investigate the ensuing modifications to the Lamb shift and to the electrostatic forces and comparing our results with the data from the muonic hydrogen anomaly,  we set lower bounds on the nonlocality scales. We also discuss the running of the electromagnetic coupling for the nonlocal theory.
The results obtained indicate that future experimental analyses on atomic phenomena, such as the Lamb shift, could allow to verify the presence of non-local effects on microscopic scales and  impose effective limits on the non-locality scale.

\end{abstract}

\pacs{ }
\maketitle

\section{Introduction}

Physics beyond the Standard Model (SM) of particles \cite{Georgi}-\cite{Buoninfant} has had great development in recent years, in a quest to explain phenomena such as particle mixing and oscillations, matter-antimatter asymmetry,  the dark matter and dark energy problems,   the strong CP problem, and to attempt at the formulation of a quantum theory of gravity.
Some exquisitely field theoretical phenomena, such as the anomalous magnetic moment of leptons or the Lamb Shift, offer an unparalleled sensitivity to the physics beyond the SM. On the other hand, they represent some of the most compelling and precise experimental verifications of quantum electrodynamics (QED).
The Lamb shift predicts a difference in energy between the two energy levels $2S_{1/2}$ and $2P_{1/2}$ of the hydrogen atom. Such a difference is due to radiative corrections involving photon emission/absorption, which can only be described in the context of QED.
The radiative correction caused by one-loop vacuum polarization accounts for the majority of the contribution to the Lamb shift.
Its experimental detection represented a great success of quantum field theory, being an important precision test for QED.
The phenomenon has been investigated mainly by means of pulsed laser spectroscopy applied to muonic hydrogen \cite{Blasc}-\cite{Drake}, in which a muon and a proton are joined to form the atom.
The shift is much more significant for muonic hydrogen than it is for the ordinary hydrogen atom.
This  is due to the fact that the muon is about $200$ times heavier than the electron. As a consequence the muonic hydrogen Bohr radius is correspondingly smaller than in hydrogen, and effects of the finite size of the proton on the muonic hydrogen energy levels, which affect the Lamb shift, are thus enhanced.

Another  field of research that has had a certain development in recent years is represented by the non-local theories. Such theories have originally been introduced to improve the ultraviolet behavior of the quantized theory and to resolve the ghost problem \cite{Krat}-\cite{Buoninfante}.
Besides being of general interest for quantum theories of gravity \cite{Krat}-\cite{Buoninfante2}, non-locality
also offers interesting avenues for solutions of many  problems in theoretical physics, as for example the muon
$g - 2$ anomaly  \cite{Capolupo}.
In fact, nonlocal theories and noncommutative field theories are closely related to string theory in  particle physics \cite{Blasc}-\cite{Gurau}.
Therefore, it is extremely important to test the validity of non-local theories and determine the  lower bounds on the nonlocality scales.
For these purposes, an important aid could be provided by the analysis  of non local effects on atomic systems. 
The continuous evolution of atomic physics experiments could in fact allow the detection of the corrections to phenomena of quantum electrodynamics, foreseen by non-local theories.

Given the above context, here we wish to explore some of the phenomenological consequences of a (string-inspired) nonlocal generalization of QED \cite{NLQED}. The latter is characterized by new energy scales $M_j$, in principle one for each of the basic fields involved, $j=\gamma$ and $j=e,\mu,\tau,$ etc. The other relevant scale shall be the mass of the fermions considered $m_f$. In particular, we will be interested in the low energy $|p^2|\ll m_f^2 \ll M^2_j$ and intermediate energy $ m_f^2 \ll |p^2| \ll M^2_j$ ranges.

In the low energy regime, we show that nonlocality induces a modification of the Coulomb potential. Although the latter appears already at tree-level, we perform the computation up to one-loop, analyzing the relevant generalization of the photon vacuum polarization diagram. 
The nonlocal generalization is obtained by replacing  each element of the diagram, namely vertices and propagators, with their nonlocal counterparts, which bring along exponential factors depending on the non-locality scales. In the ranges of momenta much smaller than the nonlocality scales, we make use of the perturbative expansions in the inverse of the square of the nonlocality scales.
Renormalizing the  ultraviolet divergent terms, we derive the non local form of the Coulomb potential, 
which contains four terms: the standard Coulomb one, two  nonlocal terms with no local counterpart, and 
 the nonlocal generalization of the Uehling term. In the local limit, the non local terms disappear and the Uehling  term reduces to the standard one. The expression of the non local Coulomb potential is the main result of the present work.

We then analyze the effects  of the nonlocal Coulomb potential on the energy levels of hydrogen-like atoms, focusing our attention  on muonic hydrogen, for which the non-local effects are higher. 
We derive a modification to the usual Lamb shift, in the non-relativistic regime, which could contribute to the discrepancy between the experimental
and theoretical values of energy difference between the $2 S^{F=1}_{\frac{1}{2}}$ and the $2 P^{F=2}_{\frac{3}{2}}$ 
 of muonic hydrogen. We also compare our results with the data from the muonic hydrogen anomaly in order to set lower bounds on the nonlocality scales $M_j$.
Moreover, we   generalize the standard electrostatic forces (Coulombian, Van der Waals, etc.) to the non local case.

In the intermediate energy range, we discuss the effective charge, and show how nonlocality affects the running of the electromagnetic coupling constant.
Our results could open new scenarios in the understanding of non local theories and in the determination of the non locality scale.

The paper is structured as follows. In sec. 2, we introduce the basic elements of the nonlocal theory and compute the tree level correction to the Coulomb potential. In Sec. 3 we  analyze the nonlocal vacuum polarization diagram. The bare photon propagator at one loop is renormalized in sec. 4, allowing us to derive the corresponding Coulomb potential. In sec. 5 and 6 these results are applied respectively to the Lamb shift in hydrogen-like atoms (with an emphasis on the muonic hydrogen) and to the macroscopically relevant electrostatic forces. Sec. 7 deals with the intermediate energy regime and the determination of the running coupling constant. Finally Sec. 8 is devoted to the conclusions.

\section{Tree level correction}

We wish to study the effects induced by a (string-inspired) nonlocal generalization of quantum electrodynamics \cite{NLQED} on the Coulomb potential and the observables related to it.
In this section we show that modifications arise already at tree level, due to the replacement of the photon propagator with its nonlocal generalization. We introduce the nonlocality scales $M_f$ (for fermions) and $M_{\gamma}$ (for the photon). These parameters, that control the nonlocal correction, have the dimensions of mass. In general $M_f$ may depend on the specific species considered (say electron, muon, tau, etc., therefore $f=e,\mu,\tau$); until explicitly needed, we won't specify the kind of fermion, and keep the discussion for a generic species $f$.  The nonlocal propagators read
\begin{equation}\label{NLProp1}
iG_{\mu\nu}^{NL} (p^2) = -i\frac{g_{\mu\nu}}{p^2+i\epsilon}e^{-\frac{p^2}{M^2_\gamma}}.
\end{equation}
and
\begin{equation}\label{NLProp2}
\Pi_{f, NL}(p^2)=\frac{i(\slashed{p}+m)}{p^2-m^2+i\epsilon}e^{-\frac{p^2}{M^2_f}}\,.
\end{equation}
respectively for the photon and the fermion \cite{NLQED}. Locality is recovered for $M_f, M_{\gamma} \rightarrow \infty$. Also the bare $0$-th order vertex $\gamma^{\mu}$ is replaced by the nonlocal counterpart
\begin{eqnarray}\nonumber \label{NLVert}
\Gamma^\mu_{NL}(p',p) &=&\frac{1}{2}\biggl[(p^\mu \slashed{p}'+p'^{\mu}\slashed{p}\biggl(\frac{e^{\frac{p'^2}{M^2_f}}-e^{\frac{p^2}{M^2_f}}}{p'^2-p^2}\biggr)
\\
&+& \bigl(e^{\frac{p'^2}{M^2_f}}
+e^{\frac{p^2}{M^2_f}}\bigr)\gamma^\mu\biggr]\,.
\end{eqnarray}
The equations \eqref{NLProp1}, \eqref{NLProp2} and \eqref{NLVert} provide the basic elements for setting up the Feynman diagrams of non-local electrodynamics, as done in \cite{Capolupo}. The simplest (tree-level) modification of the Coulomb potential is induced by Eq.\eqref{NLProp1}. Expanding the latter to first order in $\frac{p^2}{M_{\gamma}^2}$ one has
\begin{equation}\label{ExpNLProp1}
iG_{\mu\nu}^{NL} \simeq -i\frac{g_{\mu\nu}}{p^2}\biggl[1-\frac{p^2}{M^2_\gamma}\biggr]\,,
\end{equation}
which modifies  the Coulomb potential $V(p^2)=\frac{e^2}{p^2}$ according to
\begin{equation}\label{ppote}
\tilde{V}(p^2)=\frac{e^2}{p^2}\biggl(1-\frac{p^2}{M^2_\gamma}\biggr).
\end{equation}
The tree level correction is clearly finite, as it does not involve any internal momentum integration.
In the local limit $M_\gamma\rightarrow \infty$, we reobtain the usual Coulomb potential. Notice that the Coulomb correction at tree level due to non local effects can  already provide a lower bound  on the (photon) non-locality scale.

\section{Vacuum polarization}

\begin{figure}
  \includegraphics[width=9cm]{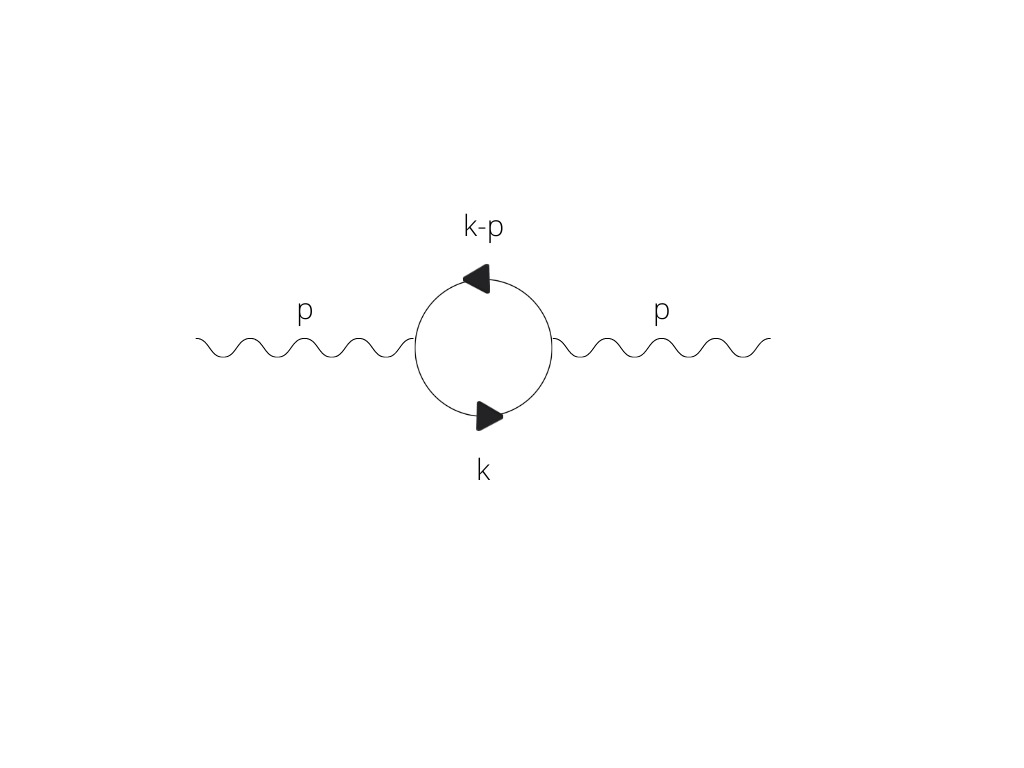}\\
\caption{Local vacuum polarization diagram}
\label{figura:vacuum}
\end{figure}

Having determined the simple tree-level correction to the Coulomb potential, we now move on to the $1$-loop (order $\alpha$) contributions. We first consider  the low momentum regime $|p^2| \ll m^2$. The basic local diagram is the vacuum polarization of Fig. \ref{figura:vacuum}.
Computation of its nonlocal generalization requires, a priori, that each element of the diagram (the two vertices, the two photon propagators and the two fermion propagators) be replaced with the nonlocal counterpart, as given in Eqs. \eqref{NLProp1}, \eqref{NLProp2} and \eqref{NLVert}. As evident from the latter, each nonlocal component brings along an exponential factor $e^{-\frac{s^2}{M^2}}$, with $s$ some combination of internal and external momenta and $M= M_{\gamma}$
or $M = M_f$. Since the non-locality scales are, for the ranges of interest to us, much larger than the (external) momenta involved, $\frac{p^2}{M^2} \ll 1$, it is meaningful to seek a perturbative expansion in the inverse of the square of the nonlocality scales $\frac{1}{M_{\gamma}^2}, \frac{1}{M_f^2}$.

\begin{figure}
\begin{center}
  \includegraphics[width=10cm]{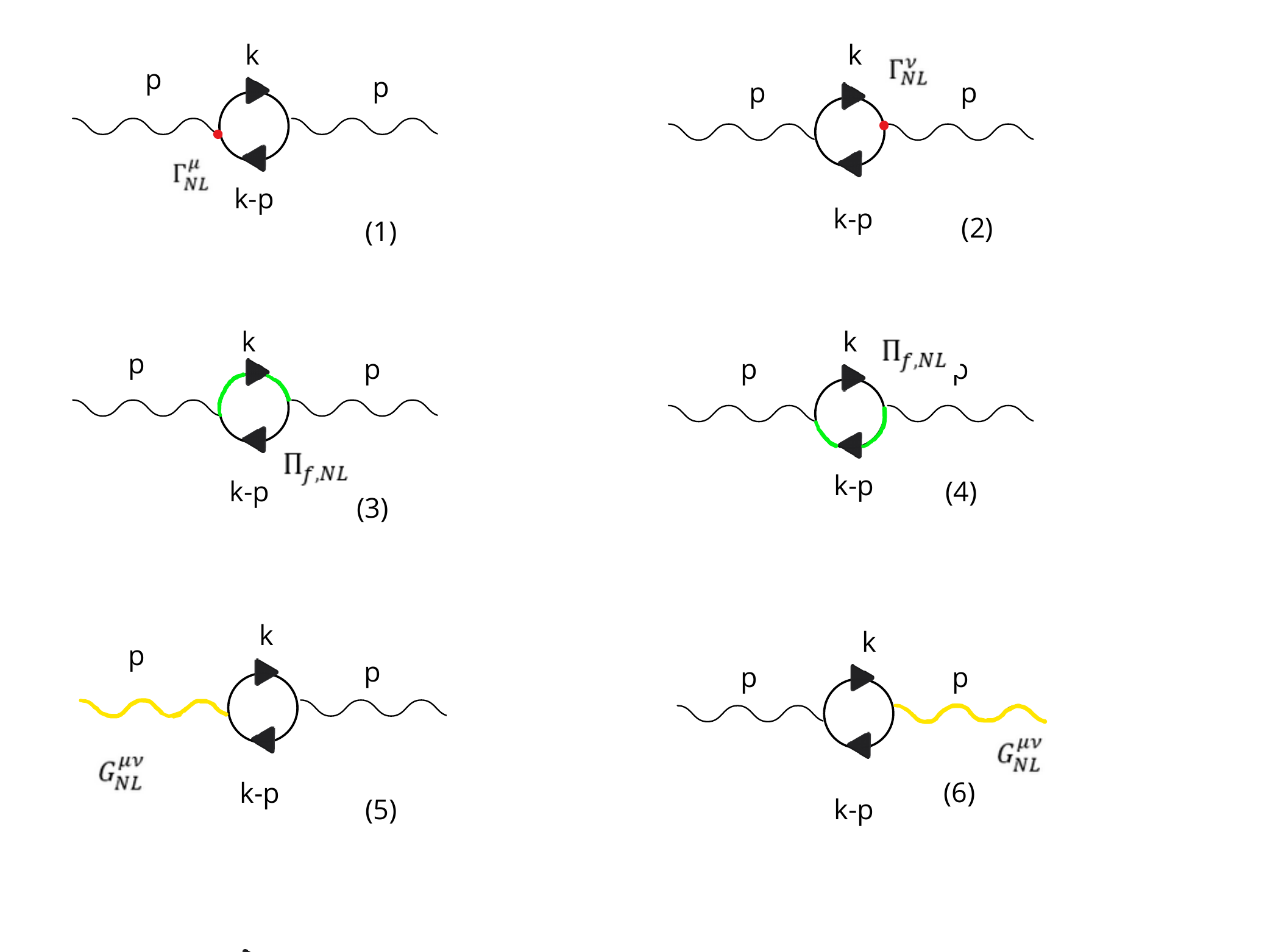}\\
\end{center}
\caption{(Color online). Relevant diagrams describing the non local vertex. Each of the local elements of the polarization diagram are replaced with their nonlocal counterpart. In the diagrams, only a single element at a time is replaced with the nonlocal counterpart. In plots (1) and (2), the vertices $\gamma^\mu$ and $\gamma^\nu$ are replaced by $\Gamma^{\mu}_{NL}$ and $\Gamma^{\nu}_{NL}$ (represented by red points in pictures), respectively. In plots (3) and (4), the fermion propagators are replaced by $\Pi_{f,NL}$ (depicted by the green lines), and in plots (5) and (6) the non local photon propagators $G^{\mu \nu}_{NL}$ (pictured   by the yellow wavy lines) substitute the standard photon propagators. }
\label{figura:vacuumNL}
\end{figure}

As discussed in \cite{Capolupo}, the contributions to the nonlocal Feynman diagram can be classified according to the order $n_{\alpha}$ in the coupling $\alpha$ and the order $m_{NL}$ in the nonlocality scales $\frac{1}{M^2}$. Denoting $(m_{NL}, n_{\alpha})$ the contributions of order $m_{NL}$ and $n_{\alpha}$, the tree-level nonlocal correction of the previous section corresponds to $(1,0) $. In this section we will deal with the order $(1,1)$ contributions. To order $m_{NL} = 1$, it is sufficient to replace each of the local elements of the polarization diagram with their nonlocal counterpart, one at a time, and sum the strictly nonlocal contributions (of order $m_{NL}=1$). Indeed, the difference between this and the diagram in which all the elements are delocalized simultaneously is only in terms of order $m_{NL}>1$. Obviously the final result up to order $(1,1)$ has to count the local limit of the vacuum polarization (of order $(0,1)$) only once. The relevant diagrams are the six shown in Fig. \ref{figura:vacuumNL}, where, in each of the diagrams, only a single element is replaced with the nonlocal counterpart.
In plots (1) and (2), the vertices  are replaced by the non local ones, in plots (3) and (4), the non local fermion propagators are considered, and in plots (5) and (6), the non local photon propagators are taken into account.

For the computation of all these diagrams it is of course sufficient to truncate the nonlocal elements to the first non-trivial order in the nonlocality scales, namely Eq. \eqref{ExpNLProp1} and
\begin{gather}
\Pi_{f,NL}\simeq \frac{i(\slashed{p}+m)}{p^2-m^2+i\epsilon}\biggl(1-\frac{p^2}{M_f^2}\biggr)\\
\Gamma^\mu_{NL}(p,p')\simeq\gamma^\mu+\frac{1}{2M_f^2}\biggl[p^\mu\slashed{p}'+p'^\mu\slashed{p}+(p'^2+p^2)\gamma^\mu\biggr]\,.
\end{gather}
Let us denote the local vacuum polarization diagram with
\begin{widetext}
\begin{eqnarray}\nonumber
 i \Pi^{\mu \nu}_{L} (p)& =& -(-ie)^2\int\frac{d^4k}{(2\pi)^4}\frac{i}{(p-k)^2-m^2+i\epsilon} \frac{i}{k^2-m^2+i\epsilon}
\mathrm{Tr}\left(\gamma^{\mu}(\slashed{k}-\slashed{p}+m)\gamma^\nu(\slashed{k}+m)\right)
\\
&=& ie^2(-p^2 g^{\mu \nu} + p^{\mu} p^{\nu})  \Pi_L (p^2).
\end{eqnarray}
While the diagrams (1) to (4) of Fig. \ref{figura:vacuumNL} modify the inner loop structure, the diagrams 5 and 6 do only modify the external photon lines, by means of the nonlocal exponential factors (see Eq. \eqref{NLProp1}). It is then straightforward to see that the (strictly) nonlocal contribution corresponding to diagrams (5) and (6), to order $m_{NL}=1$, is
\begin{equation}
 i \Pi^{\mu \nu}_{5} (p) = i \Pi^{\mu \nu}_{6} (p) =-\frac{p^2}{M_{\gamma}^2} i \Pi^{\mu \nu}_L (p) \ .
\end{equation}
The first four diagrams instead require more care. In formulae they are

\begin{equation}\label{p1}
i\Pi^{\mu\nu}_1(p)=-(-ie)^2\int\frac{d^4k}{(2\pi)^4}\frac{i}{(p-k)^2-m^2+i\epsilon} \frac{i}{k^2-m^2+i\epsilon}
\mathrm{Tr}\biggl(\Gamma^\mu_{NL}(p',k')(\slashed{k}-\slashed{p}+m)\gamma^\nu(\slashed{k}+m)\biggr),
\end{equation}
\begin{equation}\label{p2}
i\Pi^{\mu\nu}_2(p)=-(-ie)^2\int\frac{d^4k}{(2\pi)^4}\frac{i}{(p-k)^2-m^2+i\epsilon} \frac{i}{k^2-m^2+i\epsilon}
\mathrm{Tr}\biggl(\gamma^\mu(\slashed{k}-\slashed{p}+m)e^{\frac{(k-p)^2}{M_f^2}}\gamma^\nu(\slashed{k}+m)\biggr),
\end{equation}
\begin{equation}\label{p3}
i\Pi^{\mu\nu}_3(p)=-(-ie)^2\int\frac{d^4k}{(2\pi)^4}\frac{i}{(p-k)^2-m^2+i\epsilon} \frac{i}{k^2-m^2+i\epsilon}
\mathrm{Tr}\biggl(\gamma^\mu(\slashed{k}-\slashed{p}+m)\Gamma^\nu(p',k')(\slashed{k}+m)\biggr),
\end{equation}
\begin{equation}\label{p4}
i\Pi^{\mu\nu}_4(p)=-(-ie)^2\int\frac{d^4k}{(2\pi)^4}\frac{i}{(p-k)^2-m^2+i\epsilon} \frac{i}{k^2-m^2+i\epsilon}
\mathrm{Tr}\biggl(\gamma^\mu(\slashed{k}-\slashed{p}+m)\gamma^\nu(\slashed{k}+m)e^{\frac{k^2}{M_f^2}}\biggr).
\end{equation}
Since we are interested in the lowest order in $p^2/M_f^2$, by computing the traces, the integrals in Eqs.(\ref{p1})-(\ref{p4}) become:
\begin{equation}
\begin{split}\label {i1}
i\Pi^{\mu\nu}_1(p)=&\frac{2e^2}{M^2_f}\int\frac{d^4k}{(2\pi)^4}\frac{1}{[(p-k)^2-m^2+i\epsilon][k^2-m^2+i\epsilon]}\biggl\{k^\mu k^\nu[2(p-k)\cdot k-(p-k)\cdot p]+4(p-k)^\mu k^\nu[2k\cdot k-k\cdot p]+\\
&+(p\cdot k-k\cdot k+m^2)[4k^\mu(p-k)^\nu +4(k^2+(p-k)^2+4(p-k)^\mu k\nu]-4k^\mu p^\nu(p-k)\cdot k-4(p-k)^\mu p^\nu k\cdot k+\\
&+4[k^2+(p-k)^2][2k^\mu k^\nu-p^\mu k^\nu-p^\nu k^\mu]\biggr\},\\
\end{split}
\end{equation}
\begin{equation}
\label{eq: p2}
i\Pi^{\mu\nu}_2(p)=\frac{4e^2}{M^2_f}\int\frac{d^4k}{(2\pi)^4} \frac{[2k^\mu k^\nu-p^\mu k^\nu-p^\nu k^\mu+g^{\mu\nu}(p\cdot k-k\cdot k+m^2)](k-p)^2}{[(p-k)^2-m^2+i\epsilon][k^2-m^2+i\epsilon]},
\end{equation}
\begin{equation}
\begin{split}
i\Pi^{\mu\nu}_3(p)&=\frac{2e^2}{M^2_f}\int\frac{d^4k}{(2\pi)^4}\frac{1}{[(p-k)^2-m^2+i\epsilon][k^2-m^2+i\epsilon]}\biggl\{ (p-k)^\nu k^\mu(2k\cdot k-p\cdot k)+k^\nu k^\mu[2k\cdot (p-k)-p\cdot (p-k)]+\\
&+4(p\cdot k-k\cdot k+m^2)[(p-k)^\nu k^\mu+k^\nu (p-k)^\mu+\bigl((p-k)^2+k^2\bigr)g^{\mu\nu}]-4(p-k)^\nu p^\mu k\cdot k-k^\nu p^\mu(p-k)\cdot k+\\
&+4[(p-k)^2+k^2]\cdot[2k^\mu k^\nu -p^\mu k^\nu-p^\mu k^\nu]\biggr\},
\end{split}
\end{equation}
\begin{equation}\label {i4}
i\Pi^{\mu\nu}_4(p)=\frac{4e^2}{M^2_f}\int\frac{d^4k}{(2\pi)^4} \frac{[2k^\mu k^\nu-p^\mu k^\nu-p^\nu k^\mu+g^{\mu\nu}(p\cdot k-k\cdot k+m^2)]k^2}{[(p-k)^2-m^2+i\epsilon][k^2-m^2+i\epsilon]},
\end{equation}
respectively.
In the following, we consider each of the integrals (\ref{i1})-(\ref{i4}) separately. We start by computing explicitly the term $i\Pi^{\mu\nu}_2$, but a similar discussion applies to the other diagrams.

By taking into account the Lorentz invariance,  $\Pi^{\mu\nu}_2 $ can be written in the most general form as (see for instance \cite{Schwartz})
\begin{equation}\label{p-par}
\Pi^{\mu\nu}_2 =\Delta_1(p^2, m^2)p^2 g^{\mu\nu}+\Delta_2 (p^2, m^2)p^\mu p^\nu\,,
\end{equation}
with $\Delta_1  $ and $\Delta_2 $ some form factors. To understand the role of these form factors, we consider, for example,  the photon propagator corresponding to the sum of the tree level terms and $\Pi_2^{\mu \nu}$. Using the parametrization of Eq.(\ref{p-par}), it can be expressed as
\begin{eqnarray}\nonumber
 iG^{\mu\nu}(p)&=&-i\frac{g_{\mu\nu}}{p^2+i\epsilon}e^{\frac{-p^2}{M^2_\gamma}}+O(e^2)
 \\\nonumber
&=&-i\frac{g_{\mu\nu}}{p^2+i\epsilon}e^{\frac{-p^2}{M^2_\gamma}}+\frac{-ig^{\mu\alpha}}{p^2+i\epsilon}i\Pi_{2,\alpha\beta}\frac{-ig^{\beta\nu}}{p^2+i\epsilon}+O(e^4)
\\\nonumber
&=& -i\frac{g_{\mu\nu}}{p^2+i\epsilon}e^{\frac{-p^2}{M^2_\gamma}}+\frac{-i}{p^2+i\epsilon}\biggl(\Delta_1g^{\mu\nu} +
 \Delta_2\frac{p^\mu p^\nu}{p^2}\biggr)+O(e^4)
\\
&=&\frac{-i (e^{\frac{-p^2}{M^2_\gamma}}+\Delta_1)g^{\mu\nu}+\Delta_2 \frac{p^\mu p^\nu}{p^2}}{p^2+i\epsilon}.
\end{eqnarray}
Notice that the term proportional to $\Delta_2$,  which is proportional to $p^\mu p^\nu$, gives  only corrections to the gauge, therefore it can be neglected (see \cite{Schwartz}) and we consider only the terms  containing $\Delta_1$.

Returning to the integrals (\ref{i1})-(\ref{i4}), the terms containing the product $p\cdot k$ vanish for symmetry reasons. The term $k^{\mu }k^{\nu }$ produces a $p^{\mu }p^{\nu }$ piece, but also yields a $g^{\mu \nu}$ term, which is the relevant one.
Taking into account the above simplifications, the term $i\Pi^{\mu\nu}_2$ in Eq.\eqref{eq: p2} becomes:

\begin{equation}\label{pi2}
i\Pi^{\mu\nu}_2= \frac{-4e^2}{M_f^2}\int \frac{d^4k}{(2\pi)^4} \frac{[2k^\mu k^\nu+g^{\mu\nu}(-k^2+p\cdot k+m^2](k-p)^2}{[(p-k)^2-m^2+i\epsilon][k^2-m^2+i\epsilon]}\,.
\end{equation}
\end{widetext}
The denominator in Eq.(\ref{pi2}) can be simplified, as usual, through the Feynman parameter technique:
\begin{equation}
\frac{1}{AB}=\int_0^1 dx \frac{1}{[A+(B-A)x]^2}\,.
\end{equation}
In our case, by setting $A = (p-k)^2-m^2+i\epsilon$ and $B = k^2-m^2+i\epsilon$, we have:
\begin{equation}
\frac{1}{[(p-k)^2-m^2+i\epsilon][k^2-m^2+i\epsilon]}=\int_0^1 dx \frac{1}{k^2-\Delta(p^2) +i\epsilon}
\end{equation}
where
\begin{equation}
\Delta(p^2) \equiv \Delta(p^2,x) =  m^2 - p^2 x (1-x).
\end{equation}
Moreover, shifting  the 4-momentum $k^\mu$   as $k^\mu \rightarrow k^\mu+p^\mu(1-x)$,
  the numerator in Eq.(\ref{pi2})  becomes
\begin{eqnarray}\nonumber\label{num}
&&N^{\mu\nu}(x)=(2k^\mu k^\nu -g^{\mu\nu}k^2)k^2+ g^{\mu\nu}(p^2x(1-x)+m^2)k^2
\\\nonumber
&&+(2k^\mu k^\nu -g^{\mu\nu} k^2)p^2 x^2 +g^{\mu\nu}(p^2x(1-x)+m^2)p^2x^2 .
\\
\end{eqnarray}
Adopting dimensional regularization in $d$ dimensions, we can substitute
\begin{equation}
k^\mu k^\nu \rightarrow \frac{i}{d} g^{\mu\nu} k^2 .
\end{equation}
in the $d^d k$ integral.
After straightforward computations, the numerator of Eq.(\ref{num}) can be expressed as
\begin{equation}
N^{\mu\nu}(x)=N_1 (x)k^4 g^{\mu\nu} + N_2(x) k^2 g^{\mu\nu} + N_3(x) g^{\mu\nu}
\end{equation}
with
\begin{equation}
\begin{split}
&N_1=(\frac{2}{d}-1)\\
&N_2(x)=[p^2x(1-x)+m^2+(\frac{2}{d}-1)p^2x^2]\\
&N_3(x)= [p^2x(1-x)+m^2]p^2x^2 .
\end{split}
\end{equation}
Using the master formula \cite{Schwartz}
\begin{widetext}
\begin{equation}
I=\int \frac{d^dk}{(2\pi)^d} \frac{k^{2a}}{(k^2-\Delta(p^2))^b}=\frac{i}{(4\pi)^{d/2}}\frac{(-1)^{a-b}}{\Delta(p^2)^{b-a-d/2}}\frac{\Gamma(a+d/2)\Gamma(b-a-d/2)}{\Gamma(b)\Gamma(d/2)},
\end{equation}
$\Pi^{\mu\nu}_2$ can be rewritten as:
\begin{equation}
\Pi^{\mu\nu}_2=\frac{-4e^2g^{\mu\nu}\mu^{4-d}}{M_f^2(4\pi)^{d/2}}\Gamma(2-d/2)\biggl\{ \int_0^1 dx \left[N_1(x)\left(\frac{2+d}{d-2}\right)\Delta(p^2,x)^2-N_2(x)\left(\frac{d}{2-d}\right)\Delta(p^2,x)+N_3(x)\right]\frac{1}{\Delta(p^2,x)^{2-d/2}}\biggr\}.
\end{equation}
In the above equation, $\mu$ is the standard dimension-adjusting parameter needed for dimensional regularization. We then impose $d=4-\epsilon$, with $\epsilon\rightarrow 0$, obtaining:
\begin{equation}\label{pi}
\Pi^{\mu\nu}_2=\frac{e^2g^{\mu\nu}}{8\pi^2}\frac{m^4}{M_f^2}\int_0^1dx\biggl(\frac{2}{\epsilon}-
log\bigl(\frac{\Delta(p^2,x)}{\mu^2}\bigr)\biggr)-\frac{3e^2g^{\mu\nu}}{4}\frac{m^2p^2}{M_f^2}\Pi_2(p^2)+\frac{7e^2g^{\mu\nu}}{8}\frac{p^4}{M_f^2}\int_0^1 dx \biggl(\frac{2}{\epsilon}-log\bigl(\frac{\Delta(p^2,x)}{\mu^2}\bigr)\biggr) x^2(1-x)^2
\end{equation}
where explicitly
\begin{equation}
\Pi_2(p^2)=\frac{1}{2\pi^2}\int_0^1 dx\biggl(\frac{2}{\epsilon}-log\bigl(\frac{\Delta(p^2,x)}{\mu^2}\bigr)\biggr)x(1-x)\,.
\end{equation}
Since we will be first interested in the low momentum regime $|p^2| \ll m^2$, for the moment we shall neglect the third term in
Eq.(\ref{pi}), considering only the terms proportional to $p^2$ and terms proportional to $m^4$. Then, $i\Pi^{\mu\nu}_2$ can be written as (here $\Delta (p^2)$ is shorthand for $\Delta (p^2,x)$ ):
\begin{equation}
i\Pi^{\mu\nu}_2=-\frac{3}{2}e^2g^{\mu\nu}\frac{p^2m^2}{M_f^2}\Pi_2(p^2)-\frac{e^2 g_{\mu\nu}}{(4\pi)^2}\frac{2m^4}{M^2_f}\int_0^1dx\biggl[\frac{2}{\epsilon}-log\bigl(\frac{\Delta(p^2)}{\mu^2}\bigr)\biggr].
\end{equation}
With similar calculations we have
\begin{eqnarray}
i\Pi^{\mu\nu}_1 &=& -\frac{1}{4}e^2g^{\mu\nu}\frac{p^2}{M_f^2}\Pi_1(p^2)+\frac{e^2 g_{\mu\nu}}{(4\pi)^2}\frac{19m^4}{M^2_f}\int_0^1dx\biggl[\frac{2}{\epsilon}-log\bigl(\frac{\Delta(p^2)}{\mu^2}\bigr)\biggr],
\\
i\Pi^{\mu\nu}_3 &=& -\frac{1}{4}e^2g^{\mu\nu}\frac{p^2}{M_f^2}\Pi_3(p^2)+\frac{e^2 g_{\mu\nu}}{(4\pi)^2}\frac{19m^4}{M^2_f}\int_0^1dx\biggl[\frac{2}{\epsilon}-log\bigl(\frac{\Delta(p^2)}{\mu^2}\bigr)\biggr],
\\
i\Pi^{\mu\nu}_4 &=& -\frac{1}{2}e^2g^{\mu\nu}\frac{p^2}{M_f^2}\Pi_4(p^2)-\frac{e^2 g_{\mu\nu}}{(4\pi)^2}\frac{2m^4}{M^2_f}\int_0^1dx\biggl[\frac{2}{\epsilon}-log\bigl(\frac{\Delta(p^2)}{\mu^2}\bigr)\biggr],
\end{eqnarray}
with  $\Pi_1(p^2)$, $\Pi_3(p^2)$ and $\Pi_4(p^2)$  given by
\begin{equation}
\begin{split}
&\Pi_1(p^2)=\frac{1}{2\pi^2}\int_0^1 dx\biggl(\frac{2}{\epsilon}-log\biggl(\frac{\Delta(p^2)}{\mu^2}\bigr)\biggr)[x(15-13x)+\frac{1}{2}]\\
&\Pi_3(p^2)=\frac{1}{2\pi^2}\int_0^1 dx\biggl(\frac{2}{\epsilon}-log\biggl(\frac{\Delta(p^2)}{\mu^2}\bigr)\biggr)[x(15-13x)+\frac{1}{2}]\\
&\Pi_4(p^2)=\frac{1}{2\pi^2}\int_0^1 dx\biggl(\frac{2}{\epsilon}-log\biggl(\frac{\Delta(p^2)}{\mu^2}\bigr)\biggr)[5x-7x^2+2x^3] \ . \\
\end{split}
\end{equation}

Summing up all the contributions up to order $(1,1)$, as schematically depicted in Figure 3, yields the bare nonlocal photon propagator
\begin{figure}
\begin{center}
  \includegraphics[width=\linewidth]{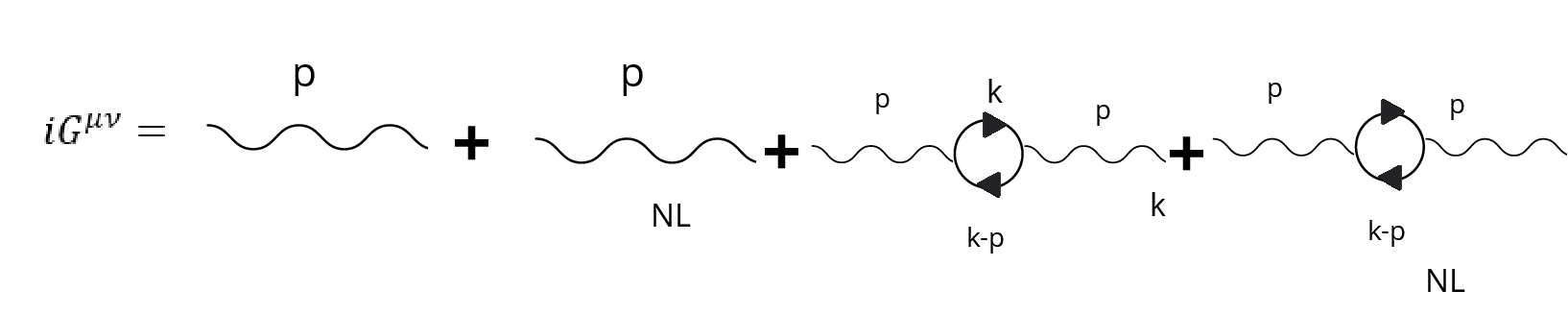}\\
\end{center}
\caption{Bare nonlocal photon propagator $iG^{\mu\nu}$, obtained by summing up all the  contributions up to order $(1,1)$.}
\label{figura:NL}
\end{figure}
\begin{equation} \label{ABBA}
\begin{split}
iG^{\mu\nu}&=-i\frac{g^{\mu\nu}}{p^2}(1-\frac{p^2}{M^2_\gamma})+\frac{-i}{p^2}i\Pi^{\mu\nu}_L\frac{-i}{p^2}+\frac{-i}{p^2}i(\Pi^{\mu\nu}_1 + \Pi^{\mu\nu}_2 +\Pi^{\mu\nu}_3+ \Pi^{\mu\nu}_4+\Pi^{\mu\nu}_5+\Pi^{\mu\nu}_6 )\frac{-i}{p^2}=\\
&-i\biggl[(1-\frac{p^2}{M^2_\gamma})
-(1-\frac{2p^2}{M^2_\gamma})e^2\Pi_L- \frac{e^2m^2}{M_f^2}\bigl(\frac{3}{2}\Pi_2 +\frac{1}{4}\Pi_1 +\frac{1}{4}\Pi_3 +\frac{1}{2}\Pi_4 \big)+\frac{34e^2m^4}{(4\pi)^2M^2_f}\int_0^1dx\biggl[\frac{2}{\epsilon}-log\biggl(\frac{\Delta(p^2)}{\mu^2}\biggr)\biggr]\biggr]\frac{g^{\mu\nu}}{p^2} \\
&= -i\left[\left(1-\frac{p^2}{M_{\gamma}^2} \right) + e^2 H(p^2) \right] \frac{g^{\mu \nu}}{p^2} \ .
\end{split}
\end{equation}
The (formally divergent) function $H(p^2)$ regroups all the terms proportional to $e^2$ and is implicitly defined by the last equality.

\section{Renormalization}

The bare photon propagator of Eq. \eqref{ABBA} contains several ultraviolet divergent terms (for $\epsilon \rightarrow 0$) and needs to be renormalized. As done in the local case, we shall tackle the divergences in Eq. \eqref{ABBA} by renormalizing the photon field strength via the factor $Z_3 = 1 + \delta_3$, where $\delta_3$ is the counterterm to be determined. The resulting counterterm has to ensure that the renormalized propagator has the correct residue at $p^2 = 0$.
It is convenient to introduce the following quantities
\begin{eqnarray}\nonumber\label{phi}
\Phi(p^2) &=& \frac{1}{2\pi^2}\int^1_0 dx\biggl[\frac{2}{\epsilon}-ln\biggl(\frac{\Delta(p^2)}{\mu^2}\biggr)\biggr]\biggl[\frac{1}{4}+\frac{23}{2}x-\frac{23}{2}x^2+x^3\biggr],
\\
\\
\theta(p^2) &=& \biggl(1-\frac{2p^2}{M^2_\gamma}\biggr))e^2\Pi_L(p^2)+\frac{e^2m^2}{M^2_f}\Phi(p^2)
\end{eqnarray}
so that the function $H(p^2)$ in Eq. \eqref{ABBA} reads
\begin{equation}\label{Hp}
H(p^2) =-\frac{\theta(p^2)}{e^2} + \frac{34 m^4}{(4\pi)^2 M_f^2} \int_0^1dx\biggl[\frac{2}{\epsilon}-log\biggl(\frac{\Delta(p^2)}{\mu^2}\biggr)\biggr] \ .
\end{equation}
The second term of the above equation is problematic. To understand what its contribution amounts to, let us expand the logarithm term in  Eq.(\ref{Hp}) in the range of interest to us ($|p^2|\ll m$):
\begin{equation}
 \frac{34 m^4}{(4\pi)^2 M_f^2} \int_0^1dx\biggl[\frac{2}{\epsilon}-log\biggl(\frac{\Delta(p^2)}{\mu^2}\biggr)\biggr] \simeq \frac{m^2}{M_f^2}\psi+\frac{m^4}{p^2M_f^2}\frac{34}{(4\pi)^2}\biggr(\frac{2}{\epsilon}+log\frac{\mu^2}{m^2}\biggl)
\end{equation}
where $\psi=\frac{34}{(4\pi)^2}\int_0^1x(1-x)$. The first term, independent of $p$, is removed when subtracting the (opportunely defined) counterterm and imposing the renormalization condition. The second modifies the pole structure of the propagator, being infrared divergent for $p^2 \rightarrow 0$. Taming this kind of divergence cannot be achieved by means of a simple counterterm, and its full treatment lies beyond the scope of this work. In the following we shall keep only the $\theta(p^2)$ term in Eq.(\ref{Hp}).
With the addition of the counterterm, the propagator becomes
\begin{equation}
iG^{\mu\nu}(p^2)=\frac{-ig^{\mu\nu}}{p^2}\biggl[\biggl(1-\frac{p^2}{M^2_\gamma}\biggr)-\biggl(1-\frac{p^2}{M^2_\gamma}\biggr)e^2\Pi_L(p^2)-e^2\frac{m^2}{M^2_f}\phi(p^2)-\delta_3\biggr] = \frac{-ig^{\mu\nu}}{p^2}\biggl[\biggl(1-\frac{p^2}{M^2_\gamma}\biggr)-\theta (p^2) - \delta_3\biggr]
\end{equation}
and the renormalization condition reads
\begin{eqnarray}\label{RenCond}
\delta_3 = - \theta(p^2=0) = - \biggl[e^2\Pi_L(0)+\frac{e^2m^2}{M^2_f}\phi(0)\biggr]\, \ .
\end{eqnarray}
 Imposing Eq. \eqref{RenCond}, the resulting propagator is then

\begin{equation}
\begin{split}
iG^{\mu\nu}(p^2)&
=\frac{-ig^{\mu\nu}}{p^2}\biggl[\biggl(1-\frac{p^2}{M^2_\gamma}\biggr) +\theta(0)-\theta(p^2)\biggr] \\
&=\frac{-ig^{\mu\nu}}{p^2}\biggl[\biggl(1-\frac{p^2}{M^2_\gamma}\biggr)+ e^2(\Pi_L(0)-\Pi_L(p^2))+\frac{e^2m^2}{M^2_f}(\phi(0)-\phi(p^2))+\frac{2p^2}{M^2_\gamma}e^2\Pi_L(p^2)\biggr].
\end{split}
\end{equation}

Notice that the last term is still divergent. Nonetheless this remaining divergence can be removed through a harmless modification of the counterterm. Indeed any counterterm of the form
\begin{equation}
\delta(p^2)=\delta_3+p^2f(p^2)
\end{equation}
with $f$ a generic function not singular in $p^2$, and $\delta_3$ satisfying condition \eqref{RenCond}, does still satisfy the same renormalization condition.
The natural choice is
\begin{equation}
\delta(p^2)=\delta_3+\frac{2p^2}{M^2_\gamma}e^2\Pi_L(0).
\end{equation}
The fully renormalized propagator becomes, in the low momentum regime
\begin{equation}
iG^{\mu\nu}(p^2)=\frac{-ig^{\mu\nu}}{p^2}\biggl[\biggl(1-\frac{p^2}{M^2_\gamma}\biggr)+e^2
\biggl(1-\frac{2p^2}{M^2_\gamma}\biggr)\biggr(\frac{-p^2}{60\pi^2m^2}\biggr)-\frac{11e^2p^2}{48\pi^2M^2_f}\biggr] \ ,
\end{equation}
where it is understood that $\Pi_L (p^2)$ and $\phi(p^2)$ have been evaluated for $|p^2|\ll m^2$.
The Fourier transform of the Coulomb potential induced by this propagator is
\begin{equation} \label{V}
\tilde{V}(p^2)=\frac{e^2}{p^2}-\frac{e^2}{M^2_\gamma}-\frac{e^4}{60\pi^2m^2}\biggl(1-\frac{2p^2}{M^2_\gamma}\biggr)-e^4\frac{11}{48\pi^2M^2_f}.
\end{equation}

Note that in Eq.(\ref{V}), the term proportional to  $p^2$ is not Fourier transformable. However, this is the byproduct of our expansion truncated to the first order in $\frac{1}{M_{\gamma}^2}$. Therefore, we momentarily restore the full exponential form of this term, and we write $\tilde{V}(p^2)$ as
\begin{equation}\label{V1}
\tilde{V}(p^2)=\frac{e^2}{p^2}-\frac{e^2}{M^2_\gamma}-\frac{e^4}{60\pi^2m^2}e^{\frac{-2p^2}{M^2_\gamma}}-\frac{11e^4}{48\pi^2M^2_\gamma} \ .
\end{equation}
At the end of the calculations, we will able  to truncate again the potential term corresponding to the exponential at the first order, consistently with the other terms.

Considering Eq.(\ref{V1}), we can now take the Fourier transform of $\tilde{V}(p^2)$. It is given by
\begin{equation} \label{potential c}
V(r)=-\frac{e^2}{4\pi r}-\frac{e^2}{M_\gamma^2}\delta^3(r)-\frac{11e^4}{48\pi^2 M^2_f}\delta^3(r)-\frac{e^4}{60\pi^2 m^2}\biggl(\frac{\pi}{2}\biggr)^2\frac{M_\gamma^3 e^{\frac{-M^2_\gamma r^2}{8}}}{(2\pi)^3}.
\end{equation}
\end{widetext}
Eq.(\ref{potential c})  has four terms: the first is the Coulomb one, the fourth represents the nonlocal generalization of the Uehling term, while the second and the third are purely nonlocal terms with no local counterpart (they disappear in the local limit). It is easy to check that the fourth term approaches the standard local (Uehling) term \cite{Schwartz} when $M_{\gamma} \rightarrow \infty$.
Eq.(\ref{potential c}) represents the central result of our work. In the rest of the paper,  we analyze many effects of the nonlocal terms of the potential $V(r)$.

\section{Nonlocal corrections to the Lamb shift}

We start by determining the impact of the nonlocal terms of Eq. \eqref{potential c} on the energy levels of hydrogen-like atoms. In particular, we shall focus on muonic hydrogen, which turns out to be more sensitive to the nonlocal corrections. We stick to the non-relativistic regime and compute the energy shift due to the non local correction in first order perturbation theory as
\begin{equation}
 \Delta E_{nlm} = \bra{\psi_{nlm}} \delta V (\pmb{r}) \ket{\psi_{nlm}} \ .
\end{equation}
Here, $\ket{\psi_{nlm}}$ is the state of the hydrogen-like atom with atomic number $Z=1$ (either electronic or muonic),  the quantum numbers $n,m,l$ label the wavefunctions of the  atom, and $\delta V(r)$ represents the deviation from the Coulomb potential.  The $\delta V(r)$ deviation, including local and nonlocal terms is given by
\begin{eqnarray}\label{MMPotential}\nonumber
 \delta V_{Local + NL} (\pmb{r}) &=& -\frac{e^2}{M_\gamma^2}\delta^3(r)-\frac{11e^4}{48\pi^2 M^2_f}\delta^3(r)
 \\
& -& \frac{e^4}{60\pi^2 m^2}\biggl(\frac{\pi}{2}\biggr)^2\frac{M_\gamma^3 e^{\frac{-M^2_\gamma r^2}{8}}}{(2\pi)^3} \ .
\end{eqnarray}
The first two terms and the last one as well contribute to the energy shift of the S states. For later convenience, we focus our attention on the $2$S state,
\begin{eqnarray}
\psi_{2S} = \frac{1}{(32 \pi^2 a_f^3)^\frac{1}{2}} \left(2 - \frac{r}{a_f} \right) e^{-\frac{r}{2a_f}},
  \end{eqnarray}
  where, $a_f = \frac{1}{\alpha m}$, is the Bohr radius. Similar expressions can be derived for all the $S$ states.
 The shift in energy is then
\begin{eqnarray}\label{NonLocalEnergyShiftS}\nonumber
\Delta E^{(NL)}_{2S} &=& \Delta E^{(Tot)}_{2S} - \Delta E^{(Loc)}_{2S}
\\
&=& -\biggl[\frac{\alpha^4m^3}{2M^2_\gamma}+\frac{11\alpha^5 m^3}{24\pi M^2_f}+\frac{4\alpha^7m^3}{15 \pi M^2_\gamma}\biggr],
\end{eqnarray}
where we have introduced $\alpha = \frac{e^2}{4 \pi}$. Here, $\Delta E^{(Loc)}_{2S}$  is the term of order zero in $1/M_{\gamma}^2$ and $\Delta E^{(NL)}_{2S}$ is first order in $1/M_{\gamma}^2$. Notice that in the derivation of Eq. \eqref{NonLocalEnergyShiftS} the contribution due to the fourth term of Eq. \eqref{potential c} has been truncated at first order in $\frac{1}{M_{\gamma}^2}$.

Given the cubic dependence on the fermion mass in Eq. \eqref{NonLocalEnergyShiftS}, we can immediately see that this correction is much more significant for the muonic hydrogen, being larger by a factor $\left(\frac{m_{\mu}}{m_e}\right)^3 \simeq 8.89 \times 10^6 $.
Since the $2P$ wavefunction has no support at the origin, the second and the third terms of $V(r) $ in Eq.\eqref{potential c} are zero and only the fourth term of $V(r)$ contributes to the nonlocal energy shift of the $2P$ level. The energy shift, to the first order in $1/M_{\gamma}^2$,  is then  given by:
 \begin{equation}\label{NonLocalEnergyShiftP}
 \Delta E^{NL}_{2P}=\frac{1}{60\pi}\frac{\alpha^7 m^3}{M^2_\gamma}.
 \end{equation}
We remark that, under the approximations employed, there is no local counterpart to the $2P$ level shift.
With $m=m_{\mu}$, the muon mass, Eqs. \eqref{NonLocalEnergyShiftS} and \eqref{NonLocalEnergyShiftP} yield the (strictly nonlocal) energy shifts of the $2S$ and $2P$ levels of muonic hydrogen.

As it turns out \cite{MuonLambAnomaly} the observed energy difference between the $2S_{\frac{1}{2}}^{F=1}$ and the $2P_{\frac{3}{2}}^{F=2}$ of muonic hydrogen is lower than its theoretical estimate. Such a discrepancy is also known as muonic hydrogen anomaly. Intriguingly, the non local corrections of Eqs. \eqref{NonLocalEnergyShiftS} and \eqref{NonLocalEnergyShiftP} are just of the right sign to shift the energy difference
\begin{equation}
 \delta E = E (2S_{\frac{1}{2}}^{F=1}) - E(2P_{\frac{3}{2}}^{F=2})
\end{equation}
downwards, by lowering $E (2S_{\frac{1}{2}}^{F=1})$ and raising $E(2P_{\frac{1}{2}}^{F=1})$. Precisely one finds the experimental and theoretical values as \cite{MuonLambAnomaly}
\begin{eqnarray*}
 \delta E_{TH} &=& -205.984 \ \mathrm{meV} \\
 \delta E_{EXP} &=& -206.295 \ \mathrm{meV} \ ,
\end{eqnarray*}
with a discrepancy of about $0.311 \ \mathrm{meV}$.

If the discrepancy is due to the non local effects, we can set a lower bound on the non-locality scale. For instance, setting $M_{\gamma} \rightarrow \infty$ for simplicity, one finds that $M_f \geq 10^{-1} \mathrm{GeV}$. In the limit $M_f \rightarrow \infty$ one finds $M_{\gamma}\geq 2.3  \ \mathrm{GeV}$. More generally we obtain a region in the $(M_{\gamma}, M_f)$ plane where the inequality $|\Delta E_{2S}^{(NL)} |\leq 0.311 \ \mathrm{meV}$
is satisfied. This region is represented in  Fig. 4.

\begin{figure}
\includegraphics[width=\linewidth]{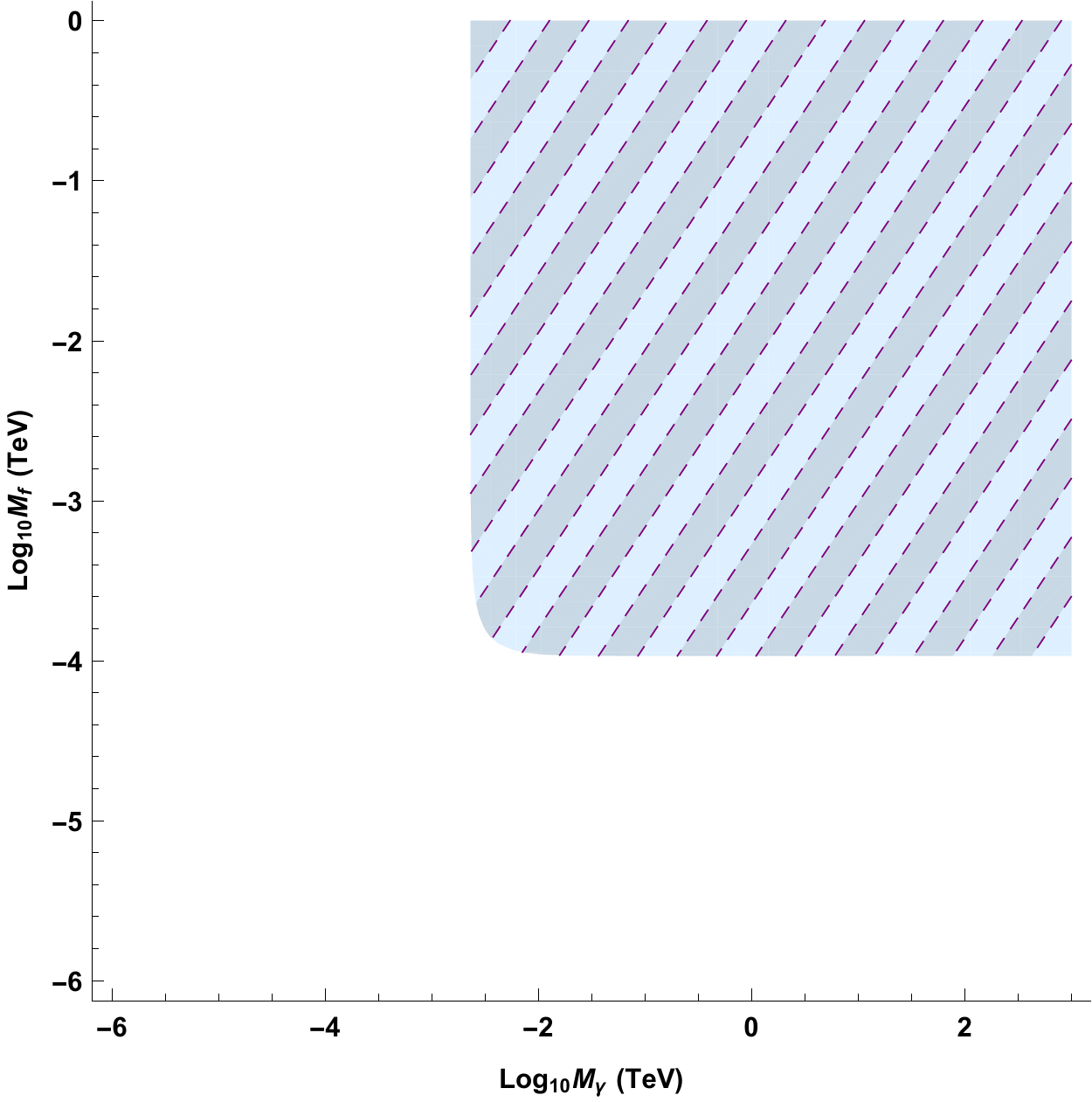}
\caption{ (Dashed area) Region of non locality scales compatible with the muonic hydrogen anomaly . The range is $(M_\gamma,M_f) \in [10^{-4},1] \times [10^{-6},10^{3}] \ \mathrm{TeV}$.}
\end{figure}

Overall we can see that the muonic hydrogen anomaly places a much milder lower bound on the non locality scales, if compared to the muon anomaly \cite{Capolupo}. As a result the muonic hydrogen anomaly might be explained, in the context of non local electrodynamics, through comparatively small non local scales. Given the bounds set by the muon anomaly \cite{Capolupo} $M_f, M_{\gamma} \geq 1 \ \mathrm{TeV}$, we conclude that while non-local electrodynamics eases the discrepancy, it cannot, alone, account for the $0.311 \ \mathrm{meV}$ difference.

\section{Nonlocal electrostatic forces}

The interaction potential of Eq. \eqref{potential c} describes the Coulomb attraction between charges of opposite sign, modified by the effect of nonlocality. For the following discussion we shall always consider a finite distance between the charges, so that the contact $\delta$ terms can be ignored. Considering the general case in which all the charged leptons (electron, muon and tau) are present, the reduced potential reads
\begin{equation}\label{RedPot}
 V(r) = -\frac{e^2}{4\pi r} -\sum_{i}\frac{e^4}{60\pi^2 m_i^2}\biggl(\frac{\pi}{2}\biggr)^2\frac{M_\gamma^3 e^{\frac{-M^2_\gamma r^2}{8}}}{(2\pi)^3} \ ,
\end{equation}
with $i=e, \mu, \tau .$
A couple of comments are in order.
First, there are in principle at least three nonlocal terms with the same structure (the last one on the right hand side), one for each charged lepton. This is because, a priori, polarization diagrams for each possible internal charged fermion must be considered. The muon and tau contributions, however, are suppressed by factors of $\frac{m_e^2}{m_{\mu,\tau}^2}$. A similar consideration holds for the other possible fermion loops (quarks), so that we shall only keep the single contribution corresponding to $m=m_e$. 
Secondly, the sign of the potential energy of Eq. \eqref{RedPot} is of course the opposite when the charges are of the same sign. The nonlocal shift in Eq. \eqref{RedPot} has an immediate impact on the electrostatic forces exerted between charged bodies.

In addition, also the electrostatic forces between neutral but polarizable bodies (Van der Waals forces) are modified by the presence of the last term. Along the standard $\frac{1}{R^6}$ term there is indeed a new strictly nonlocal contribution. To determine the distance dependence of the latter, it is convenient to follow the approach used in \cite{VdW}, and study the one-dimensional Hamiltonian model for two interacting atoms.
The atoms are thereby modeled as simple harmonic oscillators with respect to the electronic displacements $x_1,x_2$, and the interaction term is given by the Coulomb potential for the four charges
\begin{equation}
 H_1 = \frac{e^2}{4\pi}\left[\frac{1}{R} + \frac{1}{R+x_1-x_2}-\frac{1}{R+x_1}-\frac{1}{R+x_2}\right],
\end{equation}
where $R$ denotes the distance between the nuclei. It is easily shown that the $\frac{1}{R^6}$ term arises from $H_1$, diagonalizing the full Hamiltonian with the introduction of the coordinates $x_{\pm} = \frac{
x_1 \pm x_2}{\sqrt{2}}$ (see \cite{VdW} for details). We now add to $H_1$ the interaction term corresponding to the nonlocal correction
\begin{eqnarray}
\nonumber H_{NL} &=& \mathcal{B} \Bigg(e^{\frac{-M^2_\gamma R^2}{8}} + e^{\frac{-M^2_\gamma (R+x_1-x_2)^2}{8}} \\ &-& e^{\frac{-M^2_\gamma (R+x_1)^2}{8}} -e^{\frac{-M^2_\gamma (R-x_2)^2}{8}}\Bigg),
\end{eqnarray}
where $\mathcal{B} = \frac{e^4 M_{\gamma}^3}{1920 \pi^3 m_e^2}$. Extracting an $e^{\frac{-M^2_\gamma R^2}{8}}$ factor and expanding up to second order in the displacements $x_1, x_2$ (the first order vanishes), we find
\begin{equation}
 H_{NL} = \frac{\mathcal{B}M_{\gamma}^2}{4} \left( 1-\frac{M_{\gamma}^2 R^2}{4}\right) e^{\frac{-M^2_\gamma R^2}{8}} x_1 x_2 \ .
\end{equation}
Upon diagonalization, this term contributes to the $x_{\pm}$ oscillator frequencies, as it is easily seen from the identity $2x_1x_2 = x_{+}^2-x_{-}^2$. Specifically
\begin{equation}
 \omega_{\pm} = \sqrt{\omega_0^2 \mp \frac{2e^2}{4\pi m R^3} \pm \frac{\mathcal{B}M_{\gamma}^2}{4} \left( 1-\frac{M_{\gamma}^2 R^2}{4}\right) e^{\frac{-M^2_\gamma R^2}{8}}} \ .
\end{equation}
Notice that the last term is strongly suppressed by the gaussian factor. Following \cite{VdW} we compute
\begin{eqnarray*}
\nonumber && \frac{1}{2}\omega_{+} + \frac{1}{2}\omega_{-} - \omega_0 \simeq \\
&& - \frac{e^4}{32 \pi^2 \omega_0^4 R^6} + \frac{\mathcal{B}M_{\gamma}^2 e^2}{64 \pi \omega_0^4 R^3} \left( 1-\frac{M_{\gamma}^2 R^2}{4}\right) e^{\frac{-M^2_\gamma R^2}{8}} \ .
\end{eqnarray*}
The last term represents the strictly nonlocal contribution to the Van der Waals force. Notice that in the local limit $M_{\gamma} \rightarrow \infty$, one recovers the standard $\frac{1}{R^6}$ behaviour, as the second term vanishes.
The above derivation of the distance behaviour of the nonlocal term cannot provide a rigorous account of its actual size. Nevertheless it is clear that in general this term shall be several orders of magnitude below the standard Van der Waals term, approaching the latter only at very small distances. While the nonlocal Van der Waals force may have a negligible impact on a single pair of molecules, it is imaginable that it may have macroscopic, and possibly measurable, consequences on large systems.

\section{Running coupling}

In the previous sections we have dealt with the low-energy regime, establishing the modifications induced by nonlocality on the Coulomb potential. Nevertheless, given the shape of the nonlocal form factors $\propto e^{\frac{-p^2}{M^2}}$, it is clear that the nonlocal corrections get more significant as the energy is increased.

We shall now consider the limit $|p^2| \gg m^2$. In order to truncate the expansion in the nonlocality scales at the lowest non-trivial order, and compute the nonlocal corrections as before, we shall also assume that the energies involved are much lesser than the nonlocality scales $|p^2| \ll M^2$, with $M$ any of $M_{\gamma}$ and $M_f$.

The study of the extremely high energy regime $|p^2| \sim M^2$ or $|p^2| \gg M^2$ falls outside the scope of this work, and requires the evaluation of the full nonlocal vacuum polarization diagram.

In the intermediate regime $m^2 \ll |p^2| \ll M^2$, the expansion to first order in the nonlocality scale is still valid, and the contributions to the vacuum polarization are formally identical to those discussed above (see e.g. Eq. \eqref{pi}). However the only relevant terms in $\Pi_1,\Pi_2,\Pi_3,\Pi_4$ are now those scaling as $p^4$, with respect to which $m^2p^2$ and $m^4$ are negligible (see Eq. \eqref{pi}). Therefore, in this regime, the $\Pi_i$ terms are given by
\begin{widetext}
\begin{eqnarray}
i\Pi^{\mu\nu}_1 &=&-\frac{e^2 g_{\mu\nu}}{4\pi^2}\frac{p^4}{M^2_f}\int_0^1dx\biggl[\frac{2}{\epsilon}-log\bigl(\frac{\Delta(p^2)}{\mu^2}\bigr)\biggr]\biggl[\frac{1}{2}x-6x^2+9x^3-\frac{7}{2}x^4\biggr],
\\
i\Pi^{\mu\nu}_2&=& \frac{7e^2 g_{\mu\nu}}{32\pi^2}\frac{p^4}{M^2_f}\int_0^1dx\biggl[\frac{2}{\epsilon}-log\bigl(\frac{\Delta(p^2)}{\mu^2}\bigr)\biggr][x^2(x-1)^2],
\\
i\Pi^{\mu\nu}_3 &=&-\frac{e^2 g_{\mu\nu}}{4\pi^2}\frac{p^4}{M^2_f}\int_0^1dx\biggl[\frac{2}{\epsilon}-log\bigl(\frac{\Delta(p^2)}{\mu^2}\bigr)\biggr]\biggl[\frac{1}{2}x-6x^2+9x^3-\frac{7}{2}x^4\biggr],
\\
i\Pi^{\mu\nu}_4 &=& -\frac{7e^2 g_{\mu\nu}}{8\pi^2}\frac{p^4}{M^2_f}\int_0^1dx\biggl[\frac{2}{\epsilon}-log\bigl(\frac{\Delta(p^2)}{\mu^2}\bigr)\biggr][x^2(x-1)^2].
\end{eqnarray}

The resulting bare photon propagator, including the tree level and the $\Pi_5, \Pi_6$ contributions, reads

 \begin{equation}
  iG^{\mu\nu}=-\frac{ig^{\mu\nu}}{p^2}\bigg[\biggl(1-\frac{p^2}{M^2_\gamma}\biggr)-\biggl(1-\frac{2p^2}{M^2_\gamma}\biggr)e^2\Pi_L(p^2)-\frac{e^2p^2}{M^2_f}\psi(p^2)-\delta_3\biggr] \ ,
  \end{equation}
with
\begin{equation}
 \psi(p^2)=-\frac{1}{4\pi^2}\int^1_0 dx\biggl[\frac{2}{\epsilon}-log\bigl(\frac{\Delta(p^2)}{\mu^2}\bigr)\biggr]\biggl(-x+\frac{131}{8}x^2-\frac{107}{8}x^3+\frac{63}{8}x^4\biggr).
\end{equation}
Renormalization is achieved, once again, by introducing an appropriate counterterm. This is
\begin{equation}
\delta(p^2)=\delta_3+p^2\biggl(\frac{2e^2}{M^2_\gamma}\Pi_L(0)-\frac{e^2}{M^2_f}\psi(0)\biggl) \ ,
\end{equation}
where $\delta_3$ is taken to satisfy the renormalization condition \eqref{RenCond} in the intermediate energy regime $m^2 \ll |p^2| \ll M^2$.
The renormalized photon propagator then can be written as
\begin{equation}\label{RenProp}
 iG^{\mu\nu}=\frac{-ig^{\mu\nu}}{p^2}\biggl[\biggl(1-\frac{p^2}{M^2_\gamma}\biggr)+e^2\biggl(1-\frac{2p^2}{M^2_\gamma}\biggr)\Pi_L'(p^2)+\frac{e^2p^2}{M^2_f}\psi'(p^2)\biggr]
\end{equation}
\end{widetext}
where
\begin{eqnarray}
 \Pi_L'(p^2)&=& \frac{1}{12\pi^2}\ln\biggl(\frac{-p^2}{m^2}\biggr),
 \\
 \psi'(p^2)&=&\frac{517}{640\pi^2}\ln\biggl(\frac{-p^2}{m^2}\biggr) \ .
\end{eqnarray}

In the above expressions, the $x$ integral has been computed taking into account that $-p^2 \gg m^2$ (recall that $-p^2$ is positive for t-channel exchange \cite{Schwartz}) and splitting the logarithms as
$\ln\left(1-\frac{p^2}{m^2}x(1-x)\right) \simeq \ln\left(\frac{-p^2}{m^2}x(1-x)\right) =  \ln\left(\frac{-p^2}{m^2}\right) + \ln \left(x(1-x) \right)$. The $x$ dependent term produces a contribution which is easily seen to be negligible with respect to the first term, so that effectively one can replace $\ln\left(1-\frac{p^2}{m^2}x(1-x)\right) \simeq \ln\left(\frac{-p^2}{m^2}\right)$ within the integrals.
The potential corresponding to Eq.\eqref{RenProp} is then
\begin{widetext}
\begin{equation}
 \tilde{V}(p^2)=\frac{e^2}{p^2}\biggl[1-p^2\left(\frac{1}{M^2_\gamma}+\frac{e^2}{6\pi^2M^2_\gamma}\ln\left(\frac{-p^2}{m^2}\right)-\frac{517e^2}{640\pi^2M^2_f}\ln\left(\frac{-p^2}{m^2}\right)\right)+e^2\biggl(\frac{1}{12\pi^2}\ln\left(\frac{-p^2}{m^2}\right)\biggr)\biggr] \ ,
\end{equation}
where it is understood that $e$ is the (1-loop) renormalized charge. As usual this can be rewritten in terms of a momentum dependent effective charge
\begin{equation}
 \tilde{V}(p)=\frac{e^2_{eff}(\sqrt{-p^2})}{p^2}
\end{equation}
where (with $Q^2 = |p^2|$)
 \begin{equation} \label{eeff}
  e^2_{eff}(Q)=e^2\left \lbrace 1+Q^2\left[\frac{1}{M^2_\gamma}+\frac{e^2}{6\pi^2M^2_\gamma}\ln\left(\frac{Q^2}{m^2}\right)-\frac{517e^2}{640\pi^2M^2_f}\ln\left(\frac{Q^2}{m^2}\right)\right]+e^2\left(\frac{1}{12\pi^2}\ln\left(\frac{Q^2}{m^2}\right)\right)\right \rbrace
 \end{equation}

Equation \eqref{eeff} can be seen as an effective charge in nonlocal QED for the intermediate energy range $m^2 \ll Q^2 \ll M^2$. In addition to the standard logarithmic term, nonlocality induces new terms scaling as $Q^2$ and $Q^2 \ln Q^2$. Considered that by hypothesis $Q^2 \ll M^2$, these corrections are always subleading with respect to the local term. Of course in the local limit $M_\gamma,M_f\rightarrow \infty$ the usual running coupling is restored
\begin{equation}
 e^2_{eff} (Q)=e^2\biggl(1+\frac{e^2}{12\pi^2}\ln\frac{Q^2}{m^2}\biggr) \ .
\end{equation}
Eq. \eqref{eeff} can be immediately rephrased in term of the running fine structure coupling

 \begin{equation}\label{alfa}
  \alpha_{eff}(Q)=\alpha \left[1+Q^2\left(\frac{1}{M^2_\gamma}+\frac{2\alpha}{3\pi M^2_\gamma}\ln\left(\frac{Q^2}{m^2}\right)-\frac{517\alpha}{160\pi M^2_f}\ln\left(\frac{Q^2}{m^2}\right)\right)+\frac{\alpha}{3\pi}\ln\left(\frac{Q^2}{m^2}\right)\right] \ .
 \end{equation}
\end{widetext}
To get a grasp of how nonlocality affects the running of the fine structure constant, we have plotted in Fig. 5 the percentage variation on $\alpha$ due to the nonlocal term  $\alpha_{NL}(Q)$, given by the $Q^2$ term of Eq. \eqref{alfa}:
\begin{equation*}\alpha Q^2\left(\frac{1}{M^2_\gamma}+\frac{2\alpha}{3\pi M^2_\gamma}\ln\left(\frac{Q^2}{m^2}\right)-\frac{517\alpha}{160\pi M^2_f}\ln\left(\frac{Q^2}{m^2}\right)\right) \ .
\end{equation*}
\begin{figure}
\includegraphics[width=\linewidth]{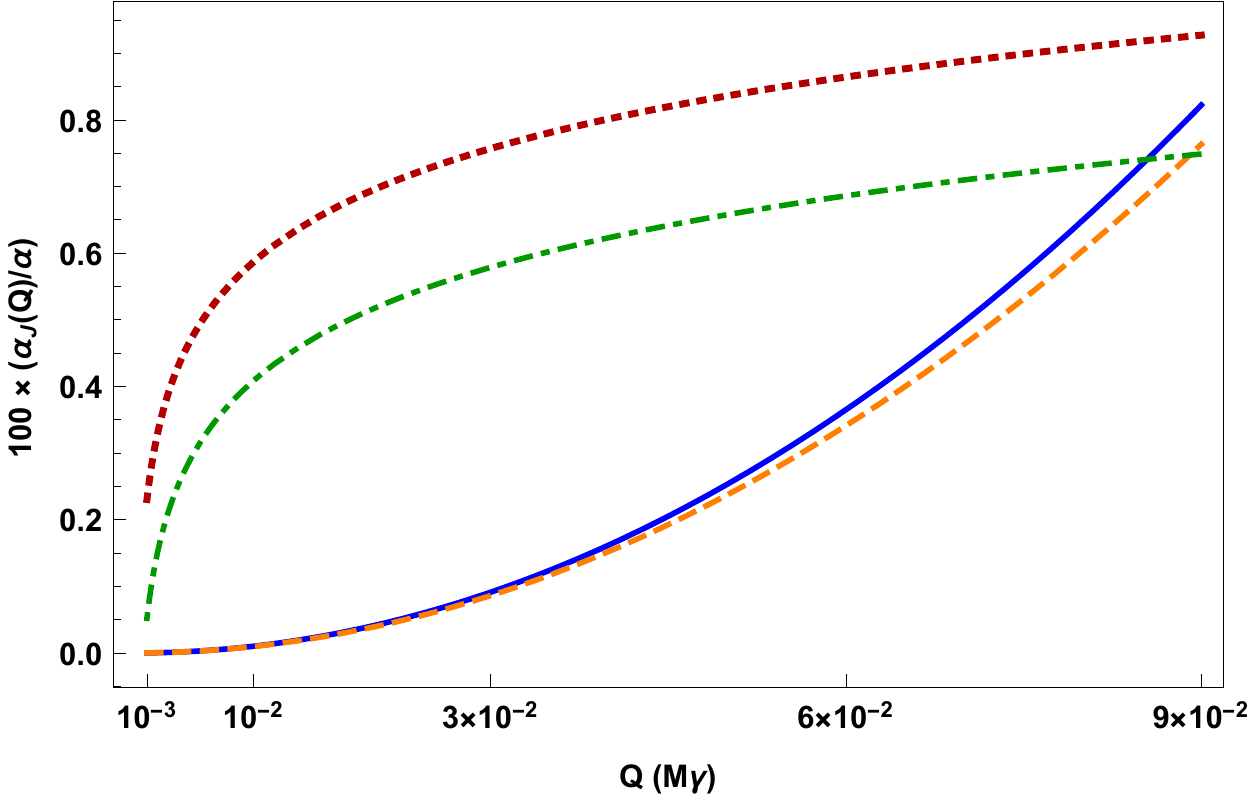}
\caption{ Percent variation of $\alpha$ due to the nonlocal correction $\alpha_J (Q) = \alpha_{NL}(Q)$ as a function of the exchanged momentum $Q$ ($Q^2=-p^2$) in the range $Q \in [0,9 \times 10^{-2}] M_{\gamma}$. (Blue solid line) For $M_f \rightarrow \infty$ and (Orange dashed line) for $M_f \simeq M_{\gamma}$. For comparison also the local correction $\alpha_J (Q) = \alpha_L (Q) = \frac{\alpha}{3\pi}\ln\left(\frac{Q^2}{m^2}\right)$ is plotted for $M_{\gamma} = 1 \ \mathrm{TeV}$ (Green dotdashed line) and $M_{\gamma} = 10 \ \mathrm{TeV}$ (Red dotted line). Notice that the precise value of $M_{\gamma}$ is essentially irrelevant for the shape of the nonlocal plots, while it is significant for the relative size of the latter with respect to the local correction. For simplicity only the electron contribution $m=m_e$ is considered.}
\end{figure}
We conclude this section spending a couple of words about the ultra-high energy regime $Q^2 \geq M^2$. Despite its limitations, Eq. \eqref{alfa} already hints at possibly large nonlocal corrections to the running coupling as the nonlocality scale is approached. Eventually, the strong energy dependency of the nonlocal corrections  (for instance $\propto Q^2$ as opposed to $\ln Q^2$ in Eq. \eqref{alfa}), might lower the energy scale of the Landau Pole of QED, determining a breakdown of the perturbation theory at scales that may be well below the $10^{286} \mathrm{eV}$ of the local theory. The complete analysis of this aspect shall be pursued in future works.

\section{conclusions}

We have analyzed some of the phenomenological aspects of a  nonlocal generalization of QED inspired by the string theory. 
In particular, considering the non local energy scales $M_j$ featuring the nonlocality and associated to  the basic fields of masses $m_f$ implied in QED, we studied the low energy $|p^2|\ll m_f^2 \ll M^2_j$ and intermediate energy $ m_f^2 \ll |p^2| \ll M^2_j$ regimes.

In the low energy regime, we computed, up to one-loop, the modification of the Coulomb potential induced by nonlocality. 
  We have shown that this  modification induces changes  to the usual Lamb shift to the   electrostatic forces such as Coulombian, and Van der Waals interactions. 
Comparing our results with the data from the muonic hydrogen anomaly, we set  lower bounds on the nonlocality scales $M_j$.
  
In the intermediate energy range, we shown that the nonlocality affects the running of the electromagnetic coupling constant.
Our results indicate that future experiments on   the Lamb shift, or Van der Waals forces
could open new ways in the study  of non-local effects, and could allow to impose effective
limits on the non-locality scale.

\section*{Acknowledgements}

We acknowledge partial financial support from MIUR and INFN. A.C. also acknowledges  the COST Action CA1511 Cosmology
and Astrophysics Network for Theoretical Advances and Training Actions (CANTATA).


\begin{thebibliography}{99}
\bibitem{Georgi}
H. Georgi and S. L. Glashow, Phys. Rev. Lett. {\bf 32}, pp.438-441 (1974); J. Wess and B. Zumino, Nucl. Phys. N {\bf 70}, pp 39-50 (1974)

\bibitem{Bilenky}
S.M. Bilenky and B. Pontecorvo, Phys. Rep. {\bf 41.4}, pp 225-261(1978); S. M. Nilenky and S. T. Petcov, Rev. Mod. Phys. {\bf 59}, pp 671-754 (1987)

\bibitem{Capolupo}
M. Blasone, A. Capolupo, G. Vitiello, Phys. Rev. D {\bf 66}, 025033 (2002); A. Capolupo, G. Lambiase, A. Quaranta, Phys. Rev. D {\bf 101}, 095022 (2020), A. Capolupo, S. M. Giampaolo, G. Lambiase, A. Quaranta, Eur. Phys. J.C.{\bf 80}, 423 (2020); A. Capolupo, S. M. Giampaolo, A. Quaranta, Phys. Lett. B {\bf 820}, 136489 (2021)

\bibitem{Ellis}
J. Ellis, Nucl. Phys. A {\bf 827.1}(2009); P. Salucci et al., Front. Phys. {\bf 8} (2021)
\bibitem{Wil}
F. Wilczek, Phys. Rev. Lett. {\bf 40}, pp. 279-282 (1978); R. D. Pec- cei and H. R. Quinn, Phys. Rev. Lett. {\bf 38}, pp. 1440-1443 (1977); G. Raffelt and L. Stodolsky, Phys. Rev. D {\bf 37},pp. 1237-1249 (1988); A. Capolupo, G. Lambiase, A. Quaranta, S. M. Gi- ampaolo Phys. Lett. B {\bf 804}, 135407 (2020); A. Capolupo, S. M. Giampaolo, A. Quaranta, Eur. Phys. J. C {\bf 81}, 1116 (2021)

\bibitem{Buoninfant}
L. Buoninfante, A. Capolupo, S. M. Giampaolo, G. Lambiase, Eur. Phys. J. C {\bf 80},  1009 (2020); A. Capolupo, S. M. Giampaolo, A. Quaranta, Eur. Phys. J. C {\bf 81},  410 (2021); K. Simonov, A. Capolupo, S. M. Giampaolo, Eur. Phys. J. C {\bf 79} 902 (2019)

\bibitem{Blasc}
D.N. Blaschke, T. Garschall, F. Gieres, F. Heindl, M. Schweda,
M. Wohlgenannt, Eur. Phys. J. C 73, 2262 (2013).

\bibitem{Giacomo}
A. Giacomo, Nucl. Phys. B {\bf 11}, 411 (1969).

\bibitem{Borie}
E. Borie and G.A. Rinker, Rev. Mod. Phys. {\bf 54}, 67 (1982)

\bibitem{Drake}
G.W.F. Drake and L.L. Byer, Phys. Rev. A {\bf 32}, 713 (1985)



\bibitem{Krat}
N. V. Krasnikov, Theor Math. Phys. 73 1184, 1987, Teor. Mat.
Fiz. 73, 235 (1987).

\bibitem{Kuz}
Yu. V. Kuzmin, Yad. Fiz. 50, 1630-1635 (1989).

\bibitem{Tomboulis}
T. Tomboulis, hep-th/9702146.

\bibitem{Bisw}
T. Biswas, A. Mazumdar and W. Siegel, JCAP 0603, 009
(2006).

\bibitem{Modesto}
L. Modesto, Phys. Rev. D 86, 044005 (2012).

\bibitem{Bisww}
T. Biswas, E. Gerwick, T. Koivisto and A. Mazumdar, Phys.
Rev. Lett. 108, 031101 (2012).

\bibitem{Biswww}
T. Biswas, A. S. Koshelev and A. Mazumdar, Phys. Rev. D 95,
043533 (2017).

\bibitem{Buoninfante}
L. Buoninfante, A. S. Koshelev, G. Lambiase and A. Mazum- dar, JCAP 1809 no.09, 034 (2018). L. Buoninfante, A. S. Koshelev, G. Lambiase, J. Marto and A. Mazumdar, JCAP 1806, no. 06, 014 (2018).

\bibitem{Buoninfante2}
L. Buoninfante, A. S. Cornell, G. Harmsen, A. S. Koshelev, G. Lambiase, J. Marto and A. Mazumdar, Phys. Rev. D 98, no. 8, 084041 (2018). V. P. Frolov, Phys. Rev. Lett. 115 (2015) no.5, 051102. S. Dengiz, E. Kilicarslan, I. Kolar and A. Mazumdar, Phys. Rev. D 102 (2020) no.4, 044016.

\bibitem{Capolupo}
 A. Capolupo, G. Lambiase, A. Quaranta, Phys. Lett. B
{\bf 829}, 137128 (2022)

\bibitem{Kato}
M. Kato, Phys. Lett. B {\bf 245}, 43 (1990)


\bibitem{NLQED}
T. Biswas and N. Okada, Nucl. Phys. B \textbf{898}, 113-131 (2015).




\bibitem{Calcagni}
G. Calcagni, L. Modesto, J. Phys. A: Math. Theor. 47 (2014)
355402.

\bibitem{Witten}
E. Witten, Nucl. Phys. B {\bf 268} (1986), 253. D.A. Eliezer, R.P.
Woodard, Nucl. Phys. B {\bf 325}, 389 (1989)

\bibitem{Koste}
 V. A. Kostelecky, S. Samuel, Nucl. Phys. B {\bf 336} (1990), 263

\bibitem{Koste2}
V. A. Kostelecky, S. Samuel, Phys. Lett. B {\bf 207} (1988), 169

\bibitem{Freund}
P. G. Freund, M. Olson, Phys. Lett. B {\bf 199} (1987), 186.

\bibitem{Freund2}
P. G. Freund, E. Witten, Phys. Lett. B {\bf 199} (1987), 191.

\bibitem{Brekke}
L. Brekke, P. G. Freund, M. Olson, E. Witten, Nucl. Phys. B
{\bf 302} (1988), 365.

\bibitem{Fra}
P. H. Frampton, Y. Okada, Phys. Rev. D {\bf 37} (1988), 3077-3079

\bibitem{Drago}
B. Dragovich, Zeta Strings, arXiv:hep-th/0703008 (2007)

\bibitem{Dou}
M. R. Douglas, S. H. Shenker, Nucl. Phys. B {\bf 335} (1990), 635

\bibitem{Gross}
D. J. Gross, A. A. Migdal, Phys. Rev. Lett. {\bf 64} (1990), 717

\bibitem{Bre}
E. Brezin, V. Kazakov, Phys. Lett. B {\bf 236} (1990), 144-150

\bibitem{Ghos}
D. Ghoshal, Phys. Rev. Lett. {\bf 97} (2006), 151601.

\bibitem{Gurau}
 R. Gurau, J. Magnen, V. Rivasseau, A. Tanasa, Commun. Math.
Phys. {\bf 287}, 275 (2009)

\bibitem{Schwartz}
M. D. Schwartz, \textit{Quantum field theory and the standard model}, Cambridge University Press (2013).

\bibitem{MuonLambAnomaly}
J. Jaeckel and S. Roy, Phys. Rev. D \textbf{82}, 125020 (2010).

\bibitem{VdW}
B. R. Holstein, American Journal of Physics \textbf{69}, pp. 441-449 (2001).
\end{thebibliography}
\end{document}